\documentclass[pra,twocolumn,showpacs,amsmath,amssymb]{revtex4}

\def\ket#1{\mathinner{|{#1}\rangle}}
\def\braket#1{\mathinner{\langle{#1}\rangle}}

\usepackage{amssymb,graphicx}
\usepackage{dsfont}
\usepackage{subfigure}

\begin{document}

\title{A measurement-induced optical Kerr interaction}
\author{Seckin Sefi}
\email{seckin.sefi@mpl.mpg.de}
\author{Vishal Vaibhav}
\author{Peter van Loock}
\email{loock@uni-mainz.de}
\affiliation{Institute of Physics, Johannes-Gutenberg Universit\"at Mainz, Staudingerweg 7, 55128 Mainz, Germany}
\affiliation{Optical Quantum Information Theory Group, Max Planck Institute for the Science of Light, G\"unther-Scharowsky-Str.1/Bau 26, 91058 Erlangen, Germany}

\begin{abstract}
We present a method for implementing a weak optical Kerr interaction (single-mode Kerr Hamiltonian)
in a measurement-based fashion using the common set of universal elementary interactions for continuous-variable quantum computation. Our scheme is a conceptually distinct alternative to the use of naturally occurring, weak Kerr nonlinearities or specially designed nonlinear media. Instead, we propose to exploit suitable offline prepared quartic ancilla states together with beam splitters, squeezers, and homodyne detectors. For perfect ancilla states and ideal operations, our decompositions for obtaining the measurement-based Kerr Hamiltonian lead to a realization with near-unit fidelity. Nonetheless, even by using only approximate ancilla states in the form of superposition states of up to four photons, high fidelities are still attainable. Our scheme requires four elementary operations
and its deterministic implementation corresponds to about ten ancilla-based gate teleportations. We test our measurement-based Kerr interaction against an ideal Kerr Hamiltonian by applying them both to weak coherent states and single-photon superposition states.
\end{abstract}

\maketitle

\section{Introduction}

Optical nonlinearities, such as the Kerr effect, have great importance in both fundamental and technological sciences. The conventional way of obtaining an optical nonlinearity is through an intensity-dependant interaction between light and matter \cite{Boyd2008}. For instance, a self-Kerr interaction naturally occurs in an optical glass fiber, inducing an intensity-dependant phase shift upon a propagating light pulse. However, such natural Kerr nonlinearities are extremely weak and a possible accumulation of the nonlinear interaction through a sufficiently long interaction time, e.g. via a long-lasting pulse propagation over a long fiber segment, is impossible due to the dominating effect of accumulating linear losses (however, see, for example \cite{SilberhornLeuchsPRL}). Thus, research on optical nonlinearities is typically based on having an exotic medium with special optical properties in order to realize a more robust, though still usually fairly weak nonlinear interaction between the medium and the light field. This research field is very active and produces various promising results \cite{Dudin2012,Peyronel2012,TanjiSuzuki2011}. Most recently, rather strong Kerr-type couplings were achieved in electro-optical circuit QED systems \cite{kirchmair}.

There is also the more recent approach of measurement-induced nonlinearities. The most prominent example for this is an efficient linear-optics quantum computation scheme that relies on entangled multi-photon ancilla states (in order to become near-deterministic) and photon counting \cite{Knill2001,Scheel2003}. This approach works on the single-photon level and uses discrete-variable qubit-type encoding. However, there are also measurement-based proposals to implement a so-called continuous-variable (CV) cubic phase gate (the hardest gate to implement from the CV universal gate set),  which, in principle, can be applied upon an arbitrary optical single-mode state \cite{Gottesman2001,Marek2011}. A very recent result is the experimental demonstration of such a (weak) cubic phase gate \cite{furusawa2012}. This particular demonstration is the key to engineer an arbitrary nonlinear interaction, since any other elementary CV gate besides the cubic phase gate is a Gaussian operation and hence fairly easy to realize \cite{Lloyd1999,Sefi2011}. However, current theoretical work on using measurement-induced, elementary nonlinear gates to obtain more useful nonlinear Hamiltonians, such as the Kerr effect, has been still insufficient with regards to an actual experimental implementation.

In this work, we present an alternative method to the medium-based Kerr nonlinearities and demonstrate obtaining the Kerr nonlinearity in a measurement-based fashion within a few measurement steps. Our implementation of the Kerr gate is independent of the input state and works, in principle, for arbitrary optical states. Interestingly, in contrast to the medium-based schemes, our approach works best on the level of individual photons. Our scheme could be more generally referred to as measurement-induced Hamiltonian engineering, however, here we shall only focus on the important example of the Kerr interaction.

The main theoretical background consists of the recently developed nonlinear-gate decomposition idea \cite{Sefi2011} combined with an approximate implementation of the cubic phase gate \cite{Marek2011}. However, besides combining them, we also extend these two ideas, presenting a rather new approach in order to avoid any redundant steps for both decomposing the Kerr interaction and preparing the ancilla states. As a result, we will be able to realize the Kerr gate using a very small number of steps (that is between 4 and 10 steps, as opposed to the tens of steps needed in the original scheme of Ref.~\cite{Sefi2011}). For implementing the elementary interactions, which are slightly more general than the common universal CV gates, we apply the measurement-based (cluster-based) model proposed in Refs.~\cite{Menicucci2006,Gu2009}. For this purpose, we first show that four elementary interactions are enough for realizing the full self(single-mode)-Kerr interaction with an interaction strength around the same magnitude as the medium-based Kerr nonlinearities. However, in order to implement these four interactions in a measurement-based and deterministic fashion using elementary CV cluster teleportations, six additional correction operations are needed. By incorporating these corrections into the measurement-based protocol, in total ten teleportations will be required.
Since the elementary operations are of cubic and quartic order in the mode operators and hence correspond to non-Gaussian operations, in the original CV cluster-based model \cite{Menicucci2006,Gu2009},
the correction operations would correspond to quadratic squeezing and further non-Gaussian operations, which must be
taken into account in the choice of the measurement bases at each step of the cluster computation.
In our ancilla-based model, however, the measurements are always homodyne detections, and for the at most cubic correction operations, suitable cubic ancilla states have to be prepared depending on the previous measurement steps.
When the approximate, heralded single-photon-based ancilla states are employed \cite{Marek2011} for this purpose, it would mean that the quantum information to be processed in the cluster has to be temporarily stored in a quantum memory \cite{Sefi2011}. Without such a memory our proposal would be probabilistic. However, if one has access to an efficient quantum memory, our weak-Kerr scheme could be also concatenated in order to implement a strong Kerr interaction.
These concatenations would accumulate the errors from each weak-Kerr scheme though, and thus,
for a strong-Kerr scheme to attain still high fidelities, the gate decompositions of the weak-Kerr schemes have to be refined through higher-order approximations, as we will show.

This work is organized as follows: in Sec.~\ref{gate_implementation}, we describe the measurement-based model of
Refs.~\cite{Menicucci2006,Gu2009}. Then, in Sec.~\ref{decomposing_the_kerr_gate}, we present a new approach to decomposing a nonlinear interaction and apply it to the Kerr interaction. After this, in Sec.~\ref{approximatingthecubic}, we will describe an experimental proposal for the necessary ancilla states and provide an explicit recipe for the implementation of the Kerr interaction. In Sec.~\ref{examples}, we test our scheme by applying the resulting decompositions to different input states and comparing the outgoing states with those obtainable via the exact evolutions. Then, in Sec.~\ref{higherorders}, we present a guideline to obtain higher-order decompositions and, finally, in Sec.~\ref{higherkerr}, we briefly discuss the possibility of a strong Kerr interaction.

Throughout, we use capital letters and hats for operators and small letters for scalars and functions. In equations, the letter $i$ is used only as the square root of -1. The notation $[A,B]$ is used for the commutation of the operators $A$ and $B$. We use the convention $\hbar=1/2$, i.e., the fundamental commutation relation is $[X,P]=i/2$ with $X\equiv (\hat{a}^\dag+\hat{a})/2$ and $P\equiv i(\hat{a}^\dag-\hat{a})/2$.

\section{Measurement-based implementation}\label{gate_implementation}

For the experimental implementation of the required operators, we use the teleportation-based model of Refs.~\cite{Menicucci2006,Gu2009}. According to this model, it is possible to teleport elementary interaction Hamiltonians (gates) that are diagonal in $X$, including rotated versions of them, onto an arbitrary input state using Gaussian coupling, homodyne detections, and an appropriate set of ancilla states.

Assume that we have the (single-mode) state $\ket{\psi}$ that we want to transform, and the operator we intend to apply is $A(X)$, which is an operator diagonal in $X$. We aim to obtain the state $\ket{\psi}_{final}=A(X)\ket{\psi}$, where
\begin{eqnarray}
\ket{\psi}&=&\int \psi(x)\ket{x}dx\,,\nonumber\\
A(X)\ket{x}&=&\alpha(x)\ket{x}\,,\nonumber\\
\ket{\psi}_{final}&=&\int \alpha(x)\psi(x)\ket{x}dx\,. \label{eq:final}
\end{eqnarray}
In order to obtain the state $\ket{\psi}_{final}$, we couple the input state $\ket{\psi}$ to an offline-prepared (single-mode) state $\ket{\alpha}$ using the Gaussian coupling operator $e^{2iX_1X_2}$, where the subscripts refer to the two modes. This coupling can be exactly implemented by beam splitters and squeezing operations \cite{braunstein2005}.
Then we make a $P$ homodyne detection on the input mode (mode 1), which yields the state $\ket{\psi}_{output}$ in mode 2 with
\begin{eqnarray}
\ket{\alpha}&=&\int \alpha(x)\ket{x}dx\label{eq:neces_ancilla}\,,\nonumber\\
\ket{\psi}_{o.p.}&=&\frac{1}{\sqrt{\pi}}\int \left(\int e^{2iy(x-\beta)} \psi(y)dy\right) \alpha(x)\ket{x}dx\,, \label{eq:output}
\end{eqnarray}
where $\beta$ is the measurement result of the homodyne detection. Equation~\eqref{eq:output} can be written as follows,
\begin{equation}
\ket{\psi}_{output}=A(X)F^{\dag}e^{2i\beta X} \ket{\psi} \,.\label{eq:output2}
\end{equation}
Here, the operator $F$ is the Fourier transform operator defined as follows,
\begin{equation}
F\ket{x}=\frac{1}{\sqrt{\pi}}\int e^{2ixy}\ket{y}dy\,,
\end{equation}
which can be used to transform the operators in $X$ to operators in $P$ and vice versa by unitary conjugation,
\begin{eqnarray}\label{eq:Fourier_transform}
Fe^{itX^m}F^{\dag}&=&e^{itP^m}\,,\nonumber\\
Fe^{itP^m}F^{\dag}&=&e^{(-1)^mitX^m}\,.
\end{eqnarray}
To obtain the desired state \eqref{eq:final} from the output state \eqref{eq:output2}, we need to apply further correction operations. For this purpose, Eq.~\eqref{eq:output2} can be written as follows,
\begin{eqnarray}
\ket{\psi}_{output}&=&A(X)F^{\dag}e^{2i\beta X} FF^{\dag}\ket{\psi}\nonumber\\
&=&A(X) e^{-2i\beta P} F^{\dag}\ket{\psi}\,.
\end{eqnarray}
Then we implement a correction operator $O_{correc}$,
\begin{eqnarray}
\ket{\psi}_{final}'&=&O_{correc} \ket{\psi}_{output}\nonumber\\
&=&A(X) F^{\dag}\ket{\psi}\,,
\end{eqnarray}
which coincides with the desired final state $\ket{\psi}_{final}$
up to the inverse Fourier transform $F^{\dag}$, and where
\begin{equation}
O_{correc}=A(X)e^{2i\beta P} A^{\dag}(X)\,.
\end{equation}
For example, when we want to implement the cubic phase gate, $A(X)=e^{itX^3}$, we need the following Gaussian correction operator,
\begin{equation}
e^{2i\beta P +3it\beta X^2}\,.
\end{equation}
For the quartic gate $A(X)=e^{itX^4}$, we need a cubic correction operator,
\begin{equation}
e^{2i\beta P +4it\beta X^3}= e^{-3it\beta^4} e^{2i\beta P} e^{4it\beta X^3}e^{-6it\beta^2 X^2}e^{4it\beta^3 X}\,.\\
\end{equation}
Note that the displacement operations, $e^{itX}$ and $e^{itP}$,
can be realized by just coupling a bright light beam with the signal state through a beam splitter.

Now in order to get rid of the inverse Fourier transform in $\ket{\psi}_{final}'$,
one possible strategy would be to first apply a Fourier transform upon the input state, $\ket{\psi}'=F\ket{\psi}$, prior to running it through the above protocol. This eventually gives the desired output state, $\ket{\psi}_{final}'=A(X) F^{\dag}\ket{\psi}'=A(X)\ket{\psi}=\ket{\psi}_{final}$.
Applying a Fourier transform in a measurement-based fashion is very easy: it just means teleporting the input state in a single, elementary step
using one of the canonical ancilla states for approximate CV cluster computation, namely a $P$ squeezed vacuum state \cite{Menicucci2006,Gu2009}.

Another strategy is to leave the inverse Fourier transform in the final states of the protocol and modify the interaction sequence in accordance to this. For example, suppose we want to teleport the four operators $A,B,C,D$ onto the initial state $\ket{\psi}$. We may then teleport the modified operators $A',B',C',D'$ onto the input and we obtain the final state,
\begin{eqnarray}\label{eq:with_fourier}
\begin{split}
A'F^{\dag}B'F^{\dag}C'F^{\dag}D'F^{\dag}\ket{\psi}\\
= A'F^{\dag}B'F^{\dag}C'F^{\dag}F^{\dag}FD'F^{\dag}\ket{\psi}\\
= A'F^{\dag}B'F^{\dag}(F^{\dag})^2(F)^2C'(F^{\dag})^2FD'F^{\dag}\ket{\psi}\\
= A'(F^{\dag})^4 (F)^3B'(F^{\dag})^3(F)^2C'(F^{\dag})^2FD'F^{\dag}\ket{\psi}\\
= A' (F)^3B'(F^{\dag})^3(F)^2C'(F^{\dag})^2FD'F^{\dag}\ket{\psi}\,,
\end{split}
\end{eqnarray}
where
\begin{eqnarray}\label{eq:ops_with_fourier}
\begin{split}
D'\equiv F^{\dag}D F,\\
C'\equiv (F^{\dag})^2C(F)^2,\\
B'\equiv (F^{\dag})^3B(F)^3,\\
A' \equiv A.
\end{split}
\end{eqnarray}
While the operators $A,B,C,D$ are assumed to be diagonal in either $X$ or $P$,
the operators $A',B',C',D'$ are straightforward to derive.

All arguments used in this section for operators diagonal in $X$ also hold in a similar way for the operators diagonal in $P$.

\section{Decomposing the Kerr interaction}\label{decomposing_the_kerr_gate}

The general problem in decomposing a given interaction is to implement the desired operator using a limited set of experimentally available operators. In our previous work \cite{Sefi2011}, we presented a systematic and efficient recipe for this, however, in order to optimize these decompositions even further with regards to an experimental implementation, here we shall use a slightly different approach. We will avoid any specific operator approximations such as splitting and (nested) commutator approximations, as employed in Ref.~\cite{Sefi2011}, and instead utilize a more brute force approach especially tailored for the Kerr interaction. This approach also has a greater flexibility for choosing the order of the operators, which turns out to be beneficial in reducing the number of experimental steps when additional correction operations are needed later on.

Let us first define the Kerr interaction as
\begin{align}
e^{it\left(X^2+P^2-\frac{1}{2}\right)^2}&=e^{it\left(X^2+P^2\right)^2}e^{-it\left(X^2+P^2-\frac{1}{4}\right)}\nonumber\\
&=e^{it\left(X^4+P^4+X^2P^2+P^2X^2\right)}e^{-it\left(X^2+P^2-\frac{1}{4}\right)}\nonumber\\
&=e^{t\left(iX^4+iP^4+\frac{4}{9}[X^3,P^3]-i\frac{1}{6}t\right)}e^{-it\left(X^2+P^2-\frac{1}{4}\right)}\nonumber\\
&\propto e^{t\left(iX^4+iP^4+\frac{4}{9}[X^3,P^3]\right)}\,.\label{eq:kerr}
\end{align}
Here, at the beginning, compared to the standard self-Kerr interaction $e^{it N(N-1)}$, with the number operator $N = X^2+P^2-\frac{1}{2}$,
we omitted the Gaussian phase-rotation operation $e^{-it N}$; and also in the last line we omitted that phase rotation and another global phase factor for simplicity. Gaussian operators are considered as relatively easy to implement and they have been already achieved experimentally \cite{PhysRevLett.106.240504}. Therefore, we define the Kerr interaction using the operators $X^4$, $P^4$, $X^3$, and $P^3$. In this form, and using decompositions into operators of this form, we can then directly apply the measurement-based model introduced in the preceding section,
Sec.~\ref{gate_implementation}. Note that in our previous work \cite{Sefi2011}, using our general and systematic decomposition scheme,
we substituted $X^4$ and $P^4$ by nested commutators of $X^3$, $P^3$, $X^2$, and $P^2$, and we employed correspondingly a universal gate set
with only quadratic and cubic gates. Here, we do allow for quartic interactions in the elementary gate set, which nonetheless can be incorporated
into the optical, ancilla-based implementation scheme.


In order to avoid splitting and commutator approximations, we aim to decompose the Kerr interaction using concatenations of the following set of interactions,
\begin{equation}\label{eq:elemantary_set}
\{e^{itX^3+itX^4},e^{itP^3+itP^4}\}\,.
\end{equation}
Taking the logarithm of arbitrary concatenations of the two elementary interactions in Eq.~\eqref{eq:elemantary_set} leads to the following elements,
\begin{eqnarray}\label{eq:lie_elements}
\begin{split}
\{tX^3,tP^3,tX^4,tP^4,t^2[X^3,P^3],\\
t^2[X^3,P^4],t^2[X^4,P^3],t^2[X^4,P^4],...\}\,.
\end{split}
\end{eqnarray}
We omit the higher-order nested commutations. Provided that $t\ll 1$, it will be sufficient for us to only deal with these elements up to a negligible error. This means that we have eight conditions to satisfy. It is generally possible to reduce the number of operators in the decomposition by reducing the number of necessary conditions to satisfy. In order to do this we divide the operators $X^3$ and $P^3$ by $\sqrt{t}$ in
Eq.~\eqref{eq:elemantary_set}. Now, the concatenations will lead to the following elements,
\begin{eqnarray}\label{eq:new_lie_elements}
\begin{split}
\{t^{1/2} X^3,t^{1/2}  P^3,t  X^4,t  P^4, t [X^3,P^3],\\
t^{3/2} [X^3,P^4], t^{3/2} [X^4,P^3], t^{2}[X^4,P^4],...\}\,.
\end{split}
\end{eqnarray}
Hence, it is sufficient to only satisfy the conditions for
\begin{equation}
\{t^{1/2} X^3,t^{1/2}  P^3,t  X^4,t  P^4, t [X^3,P^3]\}\,,
\end{equation}
neglecting any orders O$(t^{3/2})$.
Now consider the following concatenation,
\begin{equation}\label{eq:concatenation}
\begin{split}
e^{ip_1t^{1/2}P^3+ip_2tP^4}e^{ip_3t^{1/2}X^3+ip_4tX^4}\\
\times e^{ip_5t^{1/2}P^3+ip_6tP^4}e^{ip_7t^{1/2}X^3+ip_8tX^4}\,.
\end{split}
\end{equation}
Taking the logarithm of this concatenation yields
\begin{equation}\label{eq:log_concatenation}
\begin{split}
&i(p_3+p_7)t^{1/2}X^3+i(p_1+p_5)t^{1/2}P^3+i(p_4+p_8)tX^4+\\
&i(p_2+p_6)tP^4+\frac{1}{2}\left(p_1p_3-p_3p_5+p_1p_7+p_5p_7 \right)t[X^3,P^3]\\
&+{\rm O}(t^{3/2})\,.
\end{split}
\end{equation}
We then solve the corresponding polynomial equations for the coefficients $p_i$ \cite{Sefi2011} in order to obtain the logarithm of the desired Kerr operator in Eq.~\eqref{eq:kerr}. One possible solution set for Eq.~\eqref{eq:log_concatenation} is as follows,
\begin{equation}\label{eq:solution}
e^{it^{1/2}P^3}e^{\frac{4}{9}it^{1/2}X^3}e^{-it^{1/2}P^3+itP^4}e^{-i\frac{4}{9}t^{1/2}X^3+itX^4}\,.
\end{equation}
These four operators, each diagonal either in $X$ or $P$, can be implemented in a measurement-based
fashion using suitable ancilla states, as described in Sec.~\ref{gate_implementation}.
Before discussing how to possibly realize such ancilla states with quantum optics,
let us address the question whether teleporting the above four operations onto an arbitrary input state
can be done deterministically.

\subsection{With postselection}

In order to avoid the necessary correction operations when implementing the non-Gaussian cubic and quartic gates in a measurement-based fashion (recall Sec.~\ref{gate_implementation}), we may simply postselect the measurement result $\beta\approx0$, in which case each of the operators in Eq.~\eqref{eq:solution} corresponds to one teleportation step. Then we only need four non-Gaussian ancillae to decompose the Kerr interaction. Following Eqs.~\eqref{eq:with_fourier} and \eqref{eq:ops_with_fourier}, the actual operators to be teleported onto the input state are given by
\begin{eqnarray}
\begin{split}
e^{-i\frac{4}{9}t^{1/2}X^3+itX^4}\rightarrow e^{i\frac{4}{9}t^{1/2}P^3+itP^4},\\
e^{-it^{1/2}P^3+itP^4}\rightarrow e^{it^{1/2}P^3+itP^4},\\
e^{\frac{4}{9}it^{1/2}X^3}\rightarrow e^{\frac{4}{9}it^{1/2}P^3},\\
e^{it^{1/2}P^3}\rightarrow e^{it^{1/2}P^3}.
\end{split}
\end{eqnarray}

\subsection{Without postselection}

For a deterministic operation one needs to implement the corrections. In this case,
we choose to realize each cubic and quartic interaction separately,
\begin{equation}\label{eq:solution2}
e^{it^{1/2}P^3}e^{\frac{4}{9}it^{1/2}X^3}e^{-it^{1/2}P^3}e^{itP^4}e^{-i\frac{4}{9}t^{1/2}X^3}e^{itX^4}\,.
\end{equation}
This concatenation is now constructed in order to be able to implement the correction operations for the quartic interactions at the same time together with the subsequent cubic interaction and its corresponding quadratic corrections. Recall that for the quartic operator we need cubic and quadratic corrections, while for the cubic operator we need quadratic corrections. Therefore, in total we need ten elementary teleportation steps. Displacements do not require any teleportation steps, thus we do not count them.

One important subtlety, however, must be mentioned here, when the ancilla-based scheme of Sec.~\ref{gate_implementation} is to be employed
for realizing the ten teleportation steps. Since the correction operations depend on the homodyne measurement outcomes,
the corresponding ancilla states cannot all be prepared beforehand, i.e., offline in a strict sense.
For every gate that depends on a measurement outcome, the right ancilla state must be prepared effectively online.
A possible remedy for this dilemma is the use of a quantum memory, in which the quantum information to be processed
is temporarily stored until the conditional ancilla-state preparation (e.g. in the form of photonic superposition states, see later)
has succeeded \cite{Sefi2011}.

The scheme without postselection would be certainly most powerful,
if even the non-Gaussian ancilla states could be prepared in a deterministic fashion whenever they are needed.
In this case, ten deterministic teleportations with ten deterministically prepared ancillae would give a fully deterministic
Kerr gate. Compared to ten deterministic teleportations with six probabilistically prepared ancilla states (corresponding
to the six operations in Eq.~\eqref{eq:solution2} while the extra four quadratic correction operations
rely upon unconditional squeezed-state resources),
the postselected scheme with four probabilistic teleportations and four probabilistically prepared ancillae
may well be more efficient; however, the necessarily finite postselection window would also add extra errors
in addition to those caused by imperfect ancilla states.



\section{Approximating the cubic and quartic interactions}\label{approximatingthecubic}

In our measurement-based Kerr-interaction scheme, the ancilla states from Eq.~\eqref{eq:neces_ancilla} that correspond to the necessary cubic and quartic operators are
\begin{subequations}\label{eq:ancillas}
\begin{align}
\ket{\alpha}_{1}=\int e^{itx^3}\ket{x}dx\label{eq:cubic}\,,\\
\ket{\alpha}_{2}=\int e^{itx^4}\ket{x}dx\label{eq:quartic}\,,\\
\ket{\alpha}_{3}=\int e^{itx^3+it'x^4}\ket{x}dx\,.\label{eq:quartic}
\end{align}
\end{subequations}
Experimentally realizing the states in Eq.~\eqref{eq:ancillas} is the most challenging part of our scheme. For this purpose, we use the Taylor expansion approach of Ref.~\cite{Marek2011} and extend it from cubic to quartic gates (interactions). The simplest approximations for cubic and quartic interactions and their corresponding ancilla states are as follows,
\begin{subequations}\label{eq:approxs}
\begin{align}
\ket{\alpha}_{1} \approx\int (1+itx^3)\ket{x}dx\,,\label{eq:cubic_approxs}\\
\ket{\alpha}_{2}\approx \int (1+itx^4)\ket{x}dx\,,\label{eq:quartic_approxs}\\
\ket{\alpha}_{3}\approx \int (1+itx^3+it'x^4)\ket{x}dx\,.\label{eq:quartic_approxs}
\end{align}
\end{subequations}
Even though this first-order approximation is very limited, in order to obtain a small-amplitude Kerr interaction it is sufficient. Moreover, an approximate cubic interaction like in Eq.~\eqref{eq:cubic_approxs} has already been observed experimentally \cite{furusawa2012}.

The necessary ancilla states \eqref{eq:approxs} can be created by applying photon substraction and displacement operations upon a squeezed state similar to \cite{Fiurasek2005,Marek2011}. Note that a suitable squeezed vacuum state $S\ket{0}$ can be approximated by an eigenstate of the operator $P$ with eigenvalue zero, $\ket{p=0}$, such that
\begin{eqnarray}
D(-c_1)\hat{a}D(c_1)D(-c_2)\hat{a}D(c_2)D(-c_3)\hat{a}D(c_3)S\ket{0}\nonumber\\
=(\hat{a}+c_1)(\hat{a}+c_2)(\hat{a}+c_3)S\ket{0}\,,
\end{eqnarray}
corresponds to
\begin{equation}
(X+iP+c_1)(X+iP+c_2)(X+iP+c_3)\ket{p=0}\,.\label{eq:photon_substract}
\end{equation}
After expanding the state \eqref{eq:photon_substract}, we obtain the conditions for the state \eqref{eq:cubic_approxs},
namely
\begin{align*}
c_1+c_2+c_3=0\,,\\
c_1c_2+c_2c_3+c_1c_3+\frac{3}{2}=0\,,\\
itc_1c_2c_3-1=0\,.
\end{align*}
For example, for a cubic interaction with the amplitude $10^{-3}$,
the following set is one possible solution,
\begin{align*}
c_1\rightarrow 9.95i\,,\\
c_2\rightarrow -8.70356 - 4.975 i\,,\\
c_3\rightarrow 8.70356 - 4.975 i\,.
\end{align*}

The necessary procedures for the other operators can be derived in a similar fashion.

\section{Examples}\label{examples}

In this section, we apply the derived decompositions upon some test states and compare the results with those for the same states under an exact Kerr evolution. We first assume to have access to perfect ancillae. In other words, we assume that we have the elementary operators, such as $e^{itX^3+it'X^4}$, available on demand with perfect fidelity. Later we shall relax this condition and assume to have access only to approximate versions of the ancillae,
like in Eqs.~\eqref{eq:approxs}.

\subsection{Examples with ideal ancillae}\label{example}

\begin{figure}[htb!]
\centering
{
\includegraphics[width=7.5cm, height=5cm]{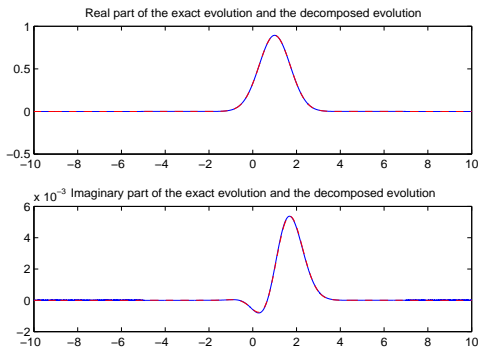}
}
\caption{Kerr evolution of a coherent state, $|\beta=1\rangle$. The position representation of the decomposed and the exact evolutions are shown. The blue line corresponds to the decomposed, the red dashed line to the exact evolution. The strength of the Kerr interaction here is $10^{-3}$.
In the corresponding measurement-based implementation, the ancillae are ideal.}
\label{fig:yeni_kerr}
\end{figure}

As an example for using the ideal ancillae, the accuracy of our Kerr decomposition is illustrated in
Fig.~\ref{fig:yeni_kerr}. The exact Kerr evolution with amplitude $10^{-3}$ and the decomposed evolution are applied to a coherent state which has an amplitude of $\beta=1$. In the position representation, this state is given by
\begin{equation}
\ket{\beta} \equiv \left(\frac{2}{\pi}\right)^{1/4}\int e^{-(x-1)^2} \ket{x}dx\,.
\end{equation}
In order to be able to calculate the exact evolution, $\sim e^{it N^2}$, one can use the well-known Fock expansion for the coherent state. Note that in Fig.~\ref{fig:yeni_kerr}, both the real and the imaginary parts of the exact and the decomposed evolutions are overlapping, indicating the high accuracy of the decomposition (while there are no additional errors, in the measurement-based model corresponding to the use of an ideal ancilla-state set).

The quality of the approximation depends on the amplitude (strength) of the applied interaction as well as on the input state. For comparison, we define an error in the decomposition through the inner product of the exactly evolved state with the output state of the same input state under the decomposed evolution: $\epsilon = 1-|\braket{\psi_{exact}|\psi_{approx.}}|$. We use the coherent state as a test input state and the error values for different configurations are listed in Table~\ref{tbl:error_configurations}. As one can see from the table, an increase of the amplitude of the coherent state also increases the error. In general, our scheme works better for weak states, in the sense that the light fields contain only a small number of photons. Similarly, an increase of the amplitude (strength) of the Kerr interaction increases the error. Nonetheless, one can also implement a reliable, high-amplitude Kerr interaction, as we will discuss later.

\begin{table}[htb!]
\begin{center}
\begin{tabular}{l|l|l}
inter. amp./ coherent state amp.  & 1 & 5  \\
 \hline
$10^{-3}$ & $10^{-7.7594}$ & $10^{-2.6641}$\\
$10^{-2}$ & $10^{-4.9653}$ & $10^{-0.7526}$\\
\end{tabular}
\end{center}
\caption{Errors for different configurations.}
\label{tbl:error_configurations}
\end{table}

Note that there are also numerical errors for obtaining the decomposed evolutions, which contribute to the total errors. It was not possible for us to extract the numerical errors alone, and so we cannot present those errors which are solely due to the interaction decomposition. In this sense, one can infer that the actual accuracy of our decompositions is even better.

\begin{figure}[htb!]
\centering
{
\includegraphics[width=7.5cm, height=5cm]{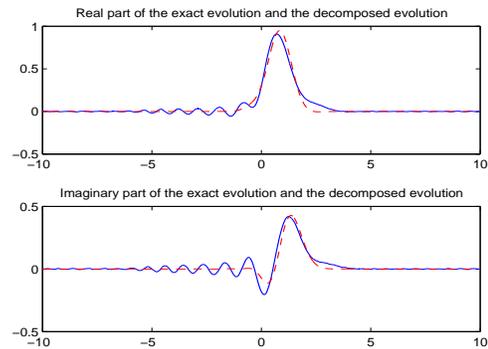}
}
\caption{Kerr evolution of a coherent state, $|\beta=1\rangle$. The position representation of the decomposed and the exact evolutions are shown. The blue line corresponds to the decomposed, the red dashed line to the exact evolution. The strength of the Kerr interaction here is $10^{-1}$.
In the corresponding measurement-based implementation, the ancillae are ideal.}
\label{fig:yeni_kerr_for_10-1}
\end{figure}

As an example for a case with a significant error, we employed our decomposition for a Kerr amplitude of $10^{-1}$, still with a coherent state where $\beta=1$. The results can be seen in Fig.~\ref{fig:yeni_kerr_for_10-1}. The errors are now clearly visible.

We also applied our decomposition upon a superposition state of a single photon and a vacuum,
\begin{equation}
\frac{1}{\sqrt{2}}(\ket{0}+\ket{1})\,,
\end{equation}
for an interaction amplitude of $10^{-3}$. This state is similar to a very weak coherent state,
$\propto \ket{0}+\beta\ket{1}$, but as opposed to a coherent state it is highly non-Gaussian. In this case, the error was $10^{-7.9853}$.

\subsection{Examples with imperfect ancillae}\label{example2}

\begin{figure}[htb!]
\centering
{
\includegraphics[width=7.5cm, height=5cm]{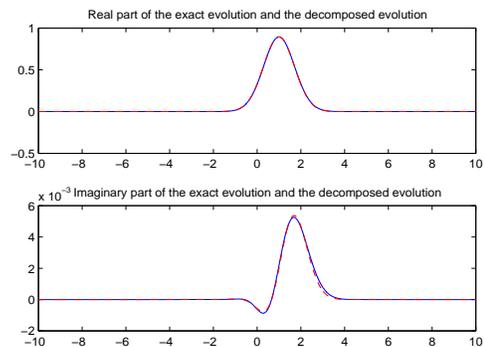}
}
\caption{Kerr evolution of a coherent state, $|\beta=1\rangle$. The position representation of the decomposed and the exact evolutions are shown. The blue line corresponds to the decomposed, the red dashed line to the exact evolution. The strength of the Kerr interaction here is $10^{-3}$. In the corresponding measurement-based implementation, the ancillae here are only approximate.}
\label{fig:yeni_approx_kerr}
\end{figure}

When all the interactions in Eq.~\eqref{eq:solution} are replaced by the operators \eqref{eq:approxs}, and again considering the previous example for the Kerr interaction with an interaction amplitude of $10^{-3}$ applied to a coherent state with amplitude 1, we obtain an evolution as shown in Fig.~\ref{fig:yeni_approx_kerr}. Compared to the case of ideal ancillae, one can see that there is an increase of the error from $10^{-7.7594}$ to $10^{-2.7446}$ due to the approximation of the ancilla states.

\section{Deriving higher-order decompositions for the Kerr interaction}\label{higherorders}

In order to be able to implement the Kerr interaction with a smaller error, we need higher-order decompositions. In this section, we will present an efficient and direct approach to derive such higher-order approximations.
Let us start with some definitions.
\begin{enumerate}

\item

Using the method explained in Sec.~\ref{decomposing_the_kerr_gate}, one can check the following operator decomposition,
\begin{eqnarray}\label{eq:supp_kerr_decomp}
\begin{split}
e^{i\frac{1}{2}t X^4} e^{itP^4+itP^3} e^{i\frac{4}{9}tX^3} e^{-itP^3} e^{i\frac{1}{2}t X^4-i\frac{4}{9}t X^3}=\\
e^{\left(itX^4+itP^4+\frac{4}{9}t^2[X^3,P^3]+t^3F_1+t^4F_2+...\right)}\,.
\end{split}
\end{eqnarray}
After the replacements $X^3\rightarrow \frac{X^3}{\sqrt{t}}$ and $P^3\rightarrow \frac{P^3}{\sqrt{t}}$, this is a second-order decomposition for the Kerr interaction and we denote it as $Q_2(t)$. For our purpose, the specific forms of the operators $F_1$, $F_2$, ... are not important.

\item

The inverse of the decomposition in Eq.~\eqref{eq:supp_kerr_decomp}, $Q_2^{-1}(t)$, can be obtained by reversing the decomposition and replacing $t$ with $-t$,
\begin{eqnarray}
 \begin{split}
e^{-i\frac{1}{2}t X^4+i\frac{4}{9}t X^3}e^{itP^3}e^{-i\frac{4}{9}tX^3}e^{-itP^4-itP^3}e^{-i\frac{1}{2}t X^4}=\\
e^{\left(-itX^4-itP^4-\frac{4}{9}t^2[X^3,P^3]-t^3F_1-t^4F_2+...\right)}\,.
\end{split}
\end{eqnarray}
This decomposition can be used for the Kerr interaction with a negative amplitude.

 \item

Thus, when we reverse the decomposition \eqref{eq:supp_kerr_decomp}, $Q_2^{rev}(t)$, we obtain the following operator,
\begin{eqnarray}
 \begin{split}
e^{i\frac{1}{2}t X^4-i\frac{4}{9}t X^3}e^{-itP^3}e^{i\frac{4}{9}tX^3}e^{itP^4+itP^3}e^{i\frac{1}{2}t X^4} =\\
e^{\left(itX^4+itP^4-\frac{4}{9}t^2[X^3,P^3]+t^3F_1-t^4F_2+...\right)}\,.
\end{split}
\end{eqnarray}

\item

Another useful decomposition is the following,
\begin{eqnarray}
 \begin{split}
e^{-i\frac{1}{2}t X^4} e^{-itP^4-itP^3} e^{-i\frac{4}{9}tX^3} e^{itP^3} e^{-i\frac{1}{2}t X^4+i\frac{4}{9}t X^3}=\\
e^{\left(-itX^4-itP^4+\frac{4}{9}t^2[X^3,P^3]-t^3F_1+t^4F_2+...\right)}\,.
\end{split}
\end{eqnarray}

%

\end{enumerate}

Now in order to derive a third-order approximation, we will use a concatenation of $Q_2(t)$ and $Q_2^{rev}(t)$. Similar approaches have been presented in Refs.~\cite{Suzuki1992b,Sefi2011}.
Hence, an order condition for a third-order approximation can be derived with the help of the following concatenation,
\begin{equation}\label{eq:third_order}
Q_2(c_1t)Q_2^{rev}(c_2t)Q_2(c_3t)Q_2^{rev}(c_4t)\,.
\end{equation}
Eliminating the third-order terms while keeping the first and second-order terms corresponds to the following order conditions,
\begin{subequations}
\begin{align}
c_1+c_2+c_3+c_4=1\,,\\
c_1^2-c_2^2+c_3^2-c_4^2=1\,,\\
c_1^3+c_2^3+c_3^3+c_4^3=0\,,\\
c_1^2 c_2 + c_1 c_2^2 + c_1^2 c_3 -c_2^2 c_3 -c_1 c_3^2 - c_2 c_3^2 +c_1^2 c_4\nonumber\\
- c_2^2 c_4+ c_3^2 c_4 + c_1 c_4^2 +c_2 c_4^2 + c_3 c_4^2=0\,.
\end{align}
\end{subequations}
The last equation corresponds to the cross terms of the first and the second-order terms.
One solution set for this equation system is the following,
\begin{eqnarray}
c_1&=&\frac{1}{6} \left(9-\sqrt{15}\right)\,,\nonumber\\
c_2&=&\frac{1}{3} \left(-3+\sqrt{15}\right)\,,\nonumber\\
c_3&=&-\sqrt{\frac{5}{3}}\,,\nonumber\\
c_4&=&\frac{1}{6} \left(3+\sqrt{15}\right)\,,
\end{eqnarray}
and by making the replacements $X^3\rightarrow \frac{X^3}{\sqrt{t}}$ and $P^3\rightarrow \frac{P^3}{\sqrt{t}}$, one can obtain a third-order decomposition for the Kerr interaction.
Further higher-order decompositions can be derived in a similar fashion.

When applying the above third-order decomposition upon a coherent state with amplitude $\beta=1$ for a Kerr interaction with amplitude $10^{-3}$, like in the example of Sec.~\ref{example}, the error in the final state decreases from $10^{-7.7594}$ to $10^{-9.9569}$.
%
An error reduction like this is crucial for concatenating weak Kerr gates (Kerr interactions with small amplitudes) sufficiently many times in order to obtain strong Kerr gates (Kerr interactions with large amplitudes), which we discuss in the next section.

\section{High-amplitude Kerr interaction}\label{higherkerr}

\begin{figure}[htb!]
\centering
{
\includegraphics[width=7.5cm, height=5cm]{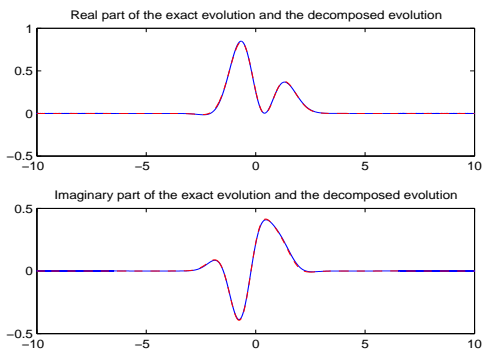}
}
\caption{
Kerr evolution of a coherent state, $|\beta=1\rangle$.
The position representation of the decomposed and the exact evolutions are shown. The blue line corresponds to the decomposed, the red dashed line to the exact evolution. A Kerr interaction with an amplitude of $10^{-3}$ is sequentially applied for 1000 times in order to get a total Kerr interaction with an amplitude 1.
In the corresponding measurement-based implementation, the ancillae are again ideal here.
}
\label{fig:high_amplitude_kerr}
\end{figure}

We can obtain a high-amplitude Kerr interaction by applying a suitable decomposition many times. As an example, we apply our third-order decomposition for 1000 times with an initial amplitude of $10^{-3}$. This enables us to obtain a Kerr amplitude of 1. We apply this whole decomposition upon a coherent state with amplitude $\beta=1$, for which the results are shown in Fig.~\ref{fig:high_amplitude_kerr}. The error for this particular decomposition is $10^{-3.8252}$.

As another example, we apply the high-amplitude Kerr gate to simulate the effect of a nonlinear sign shift gate, which can be used to entangle photons \cite{PhysRevA.52.3489}. More specifically, we apply the third-order approximation for the Kerr gate onto the following state:

\begin{equation}
\frac{1}{\sqrt{3}}(\ket{0}+\ket{1}+\ket{2}),
\end{equation}

which is supposed to transform to:

\begin{equation}
\frac{1}{\sqrt{3}}(\ket{0}+\ket{1}-\ket{2}).
\end{equation}

For the initial strength of $\pi \times 10^{-3}$, we apply the decomposition 500 times and find an error of $10^{-3.2423}$.


\section{Summary and Conclusion}

In this work, we presented an experimentally accessible method for implementing a weak Kerr interaction upon an arbitrary single-mode optical state in a measurement-based fashion. For this purpose, the Kerr interaction can be decomposed into a sequence of 4-10 elementary operations. Each Hamiltonian for these elementary operations is either cubic or quartic in $X$ and $P$ such that the necessary ancilla states in the measurement-based realization are superposition states of up to four photons. The measurements are simple, efficient homodyne detections, while the signal and ancilla states are coupled through linear beam-splitter transformations.

For the case that the homodyne measurement results are postselected around the origin in phase space,
there are no correction operations needed, and so, four elementary Hamiltonians suffice. In order to avoid such postselections, it is necessary to do the correction operations, which can be of quadratic or even cubic order
and will depend on some of the measurement outcomes. Including these corrections, ten elementary interactions are needed, and correspondingly many ancilla states. Since the cubic ancillae can be optically prepared only in a heralded fashion, in this case, quantum memories would have to be employed. If the necessary ancilla states for the elementary operations including the corrections were all available on demand in a deterministic fashion, we would obtain a fully deterministic Kerr gate.

We illustrated our Kerr decompositions with various examples, where test input states such as coherent states and single-photon states were subjected to an exact and an approximate, decomposed Kerr evolution. These examples showed that our decompositions have a high accuracy, which mainly depends on the amplitudes of the input states as well as the amplitudes of the Kerr interaction itself. The smaller these amplitudes, the higher the accuracies. This is why our scheme works best for states with fairly low (mean) photon numbers.

In order to obtain a strong Kerr interaction, we also discussed how to concatenate our decompositions for the weak Kerr interactions and showed that for this purpose, higher-order decompositions are essential in order to avoid the accumulation of errors. In principle, provided sufficiently many, suitable cubic ancilla states are available on demand, a fully deterministic, strong Kerr interaction would be implementable in a measurement-based fashion, representing an alternative approach to medium-based nonlinear dynamics.\\

\appendix*

\section{Exactly decomposing the quartic interaction}
It is also possible to exactly realize the operator $e^{itX^4}$ by Gaussian operators and cubic operators, using the two relations,
\begin{align}
e^{\frac{3}{2}it_1^2t_2X_1^2X_2}=&\,e^{-it_2X_2^3} e^{it_1 X_1P_2} e^{it_2X_2^3}e^{-2it_1 X_1P_2}  \nonumber \\
\times &e^{it_2X_2^3} e^{it_1 X_1P_2}e^{-it_2X_2^3}\,,\nonumber
\end{align}
and
\begin{equation}
e^{i\frac{t^2}{2}X_1^4} = e^{itX_1^2P_2}e^{itX_1^2X_2}e^{-itX_1^2X_2}e^{-itX_1^2P_2}\,,\nonumber
\end{equation}
where the subscripts refer to the corresponding modes. In a similar fashion, even arbitrary powers of $X$ and $P$ might be decomposed exactly.

\end{document}